\documentclass{PoS}
\usepackage{amssymb}   
\usepackage{amsmath}   
\usepackage{axodraw}

\input pix.sty

\def\undertilde#1{\mathop{\vtop{\ialign{##\cr$\textstyle{#1}$\cr%
\noalign{\kern1pt\nointerlineskip}\hfil$\mathchar"0365$\hfil\cr}}}}
\def\wideundertilde#1{\mathop{\vtop{\ialign{##\cr$\textstyle{#1}$\cr%
\noalign{\kern1pt\nointerlineskip}\hfil$\mathchar"0367$\hfil\cr}}}}

\renewcommand{\eq}{eq.~}
\renewcommand{\eqs}{eqs.~}
\renewcommand{\se}{sec.~}

\renewcommand{\fig}{fig.~}

\newcommand{\ko}{k^{ }_0}

\newcommand{\Nc}{N_{\rm c}}

\newcommand{\mE}{m_\rmii{E}}
\newcommand{\gE}{g_\rmii{E}}

\def\lsi{\raise0.3ex\hbox{$<$\kern-0.75em\raise-1.1ex\hbox{$\sim$}}}
\def\gsi{\raise0.3ex\hbox{$>$\kern-0.75em\raise-1.1ex\hbox{$\sim$}}}
\newcommand{\lsim}{\mathop{\lsi}}
\newcommand{\gsim}{\mathop{\gsi}}

\newcommand{\rmii}[1]{{\mbox{\tiny\rm{#1}}}}

\newcommand{\im}{\mathop{\mbox{Im}}}
\newcommand{\Tint}[1]{{\hbox{$\sum$}\!\!\!\!\!\!\!\int\,}_{\!\!\!\!\raise-0.9ex\hbox{$\scriptstyle{#1}$}}}
\newcommand{\Tinti}[1]{{{\Sigma}\!\!\!\!\raise0.3ex\hbox{$\int$}_\rmii{${#1}$}}}

\newcommand{\bi}{\begin{itemize}}
\newcommand{\ei}{\end{itemize}}
\newcommand{\hide}[1]{ }

\title{Towards understanding thermal jet quenching via lattice simulations}

\ShortTitle{Towards understanding thermal jet quenching 
via lattice simulations}

\author{M.\ Laine and \speaker{A.\ Rothkopf} \\
        ITP, Albert Einstein Center, University of Bern, 
        Sidlerstrasse 5, CH-3012 Bern, Switzerland\\
        E-mail: \email{laine@itp.unibe.ch}, \email{rothkopf@itp.unibe.ch}}

\abstract{
After reviewing how simulations employing classical lattice gauge theory
permit to test a conjectured Euclideanization property of a light-cone Wilson
loop in a thermal non-Abelian plasma, we show how Euclidean data can in turn be
used to estimate the transverse collision kernel, $C(k_\perp)$, characterizing
the broadening of a high-energy jet. First results, based on data produced
recently by Panero {\it et al}, suggest that $C(k_\perp)$ is enhanced over 
the known NLO result in a soft regime $k_\perp < $ a few $T$. 
The shape of $k_\perp^3 C(k_\perp)$ is consistent with a Gaussian 
at small $k_\perp$. 
}

\FullConference{31st International Symposium on Lattice Field Theory 
                LATTICE 2013\\
		July 29th - August 3rd, 2013\\
		Mainz, Germany}

\begin{document}

%
\section{Motivation}

Among the main observables measured in heavy ion collision
experiments are ``hard probes'', i.e.\ particle-like
objects having an energy much larger than the temperature. 
Hard probes can either be colour-neutral, such as photons, 
or coloured, such as jets. In the case of photons, the probe 
escapes the thermal medium unaltered, and its average production
rate reflects directly the physics of the production mechanism. 
In the case of jets, in contrast, the probe experiences 
a complicated evolution, with the jet losing
its virtuality to radiation, its energy and longitudinal momentum
to radiation and 
collisions with the medium, but simultaneously gaining transverse
momentum from collisions, leading to broadening. These 
phenomena may collectively be referred to as jet quenching; 
for reviews, see e.g.\ refs.~\cite{old}--\cite{new}. 

One quantity characterizing many of the mentioned processes
is the so-called {\rm transverse collision kernel}, denoted by 
$C(k^{ }_\perp)$. Its appearance in the context 
of jet broadening is sketched in \se\ref{se:wilson}, 
whereas a recent discussion of its role in photon production
can be found in ref.~\cite{dGamma}.
The focus of the present study is a non-perturbative estimate
of $C(k^{ }_\perp)$ with the help of lattice gauge theory, 
following ideas put forward by Caron-Huot in the context 
of an NLO computation~\cite{sch}. 

%
\section{Momentum broadening and the light-cone Wilson Loop}
\la{se:wilson}

Let $P(k^{ }_\perp,L)$ be a 
probability distribution, normalized as 
$
 \int \frac{{\rm d}^2\vec{k}^{ }_\perp}{(2\pi)^2} 
 P(k^{ }_\perp,L) = 1
$,  
of transverse momenta of a jet, once it has traversed a path 
of length $L \gg 1/\pi T$ 
within a medium of temperature $T$. The classical nature
of $P$ could originate from decoherence due to many collisions.
Energy and longitudinal momenta are
assumed hard, $\ko, k^{ }_\parallel \gg \pi T$, with  
$\pi T$ denoting a typical energy scale of a relativistic plasma, 
but virtuality is small and will be neglected in the following. 

Considering a jet seeded by a quark, $P(k^{ }_\perp,L)$ is given
by a Fourier transform of a light-cone Wilson loop
in the fundamental representation (our discussion
follows appendix~D of ref.~\cite{pheno}): 
\be
 P(x^{ }_\perp, L) \; = \;  
 \frac{1}{\Nc} \tr \langle \mathcal{W}_\rmii{F}(x^{ }_\perp, L, t) \rangle 
 \;, \quad
 P(k^{ }_\perp,L) \; = \; 
 \int \! {\rm d}^2 \vec{x}^{ }_\perp \, 
 e^{- i \vec{k}^{ }_\perp \cdot \vec{x}^{ }_\perp} \, 
 P(x^{ }_\perp, L) 
 \;, \la{P_def}
\ee
where $t$ denotes time. 
Along the light-cone, $t = L$, so normally one argument
is suppressed. 
In the case of a jet seeded by a gluon, the Wilson loop is in the
adjoint representation.
$P(x^{ }_\perp,L)$ evolves as  
\be
 \frac{{\rm d} P(x^{ }_\perp,L)}{{\rm d}L }
 = - V(x^{ }_\perp) \, P(x^{ }_\perp,L)
 \;, \la{eq_x}
\ee
where $V(x^{ }_\perp)$ 
may be called a dipole cross section (we refer to it as a transverse 
potential). The transverse collision kernel, $C(k^{ }_\perp)$, contains
the same information as $V(x^{ }_\perp)$ but in Fourier space: 
\be
 V(x_\perp) = \int \! \frac{{\rm d}^2 \vec{k}_\perp}{(2\pi)^2} 
 \Bigl( 1 - e^{i \vec{k}_\perp\cdot \vec{x}_\perp} \Bigr)
 \, C(k_\perp)
 \;. \la{relation}
\ee
Consequently, the probability distribution of the transverse momenta obeys
\be
 \frac{{\rm d} P(k^{ }_\perp,L)}{{\rm d}L }
 = 
 \int \! \frac{{\rm d}^2 \vec{q}_\perp}{(2\pi)^2}
 \; C(q^{ }_\perp) \; 
 [P(k^{ }_\perp - q^{ }_\perp,L) - P(k_\perp, L)]
 \;. \la{eq_k}
\ee
In the following we refer to the extent of the Wilson loop 
by $t$ rather than $L$.

The goal is thus to extract the damping rate of a real-time Wilson
loop, \eq\nr{eq_x}. A direct determination with Euclidean lattice QCD is 
probably beyond reach because two analytic 
continuations are needed~\cite{own}.
On the other hand it has been argued~\cite{sch}
that for $k^{ }_\perp \ll \pi T$
the dominant contribution to the collision kernel resides in 
soft thermal 
modes, which are not sensitive to the velocity of the seeding
parton. A possible strategy is to tilt the
seeding parton beyond the light cone into the space-like domain upon which the
associated Wilson loop becomes amenable to Euclidean methods.
As a first step, we test this argument by making use of classical
lattice gauge theory, which correctly represents the physics of 
the soft gauge fields~\cite{oldclas,bms}. 
A great advantage is that classical simulations
are carried out directly in real time, avoiding any analytic continuation. 

%
\section{Classical lattice gauge theory}

The Hamiltonian formalism \cite{ks} of classical lattice gauge
theory is naturally formulated in a fixed temporal gauge $A^0=0$.
Then its degrees of freedom, colour-electric fields $E_j=E^a_jT^a\in$ su(3)
and spatial links $U_j\in$ SU(3), live on three-dimensional time slices. 
The theory is parametrized by a single dimensionless number 
$\beta_\rmii{G} \equiv 2\Nc/(g^2Ta)$ with
$\Nc$ the number of colours, $g^2 \equiv 4\pi \alpha_s$ 
a renormalized gauge coupling, and
$a$ the lattice spacing, respectively. The Hamiltonian reads
\begin{align}
  H_\rmi{cl}=\sum_{\bf x} 
  \Big\{ \sum_{i=1}^3 \tr[E_i^2({\bf x})] + 
  \frac{1}{2\Nc} \sum_{i,j=1}^3 \tr[1-P_{ij}({\bf x})]\Big\}
  \;, \la{Hcl}
\end{align}
where $P_{ij}$ denotes a plaquette in the $(i,j)$-plane.
The associated local Gauss constraint 
$
  G({\bf x})=\sum_{i} 
  [ E_i({\bf x}) - U^\dagger_i({\bf x}-\hat{i})E_i({\bf x}-\hat{i})
  U_i({\bf x} - \hat{i}) ]
$
singles out the physically admissible $G({\bf x})=0$ configurations. 
The classical equations of motion read
\begin{align}
 &a\partial_tU_j({\bf x},t)
 = i(2\Nc)^{\frac{1}{2}}E_j({\bf x},t)U_j({\bf x},t) \;, \la{eom1} \\
 &a\partial_t E_i^b({\bf x},t)=-\Big(\frac{2}{\Nc}\Big)^{\frac{1}{2}}
 {\im\tr}\Big[T^bU_i({\bf x},t)\sum_{|j|\neq i}
 S^\dagger_{ij}({\bf x},t)\Big]
 \;, \label{eom2}
\end{align}
where $S_{ij}$ denotes the staple in a $(i,j)$-plane which 
closes to a plaquette when multiplied by $U_i^\dagger$.

\begin{figure}[t]


\centerline{%
  \begin{picture}(200,100)(0,0)
  \SetWidth{1.0}
  \LongArrow(70,40)(115,10)%
  \LongArrow(70,40)(70,100)%
  \DashLine(70,40)(170,55){4}%
  \LongArrow(140,50.5)(170,55)%
  \SetWidth{2.0}
  \DashLine(100,20)(115,35){1}%
  \DashLine(135,55)(150,70){1}%
  \Line(135,38)(135,53)%
  \Line(117,35)(135,38)%
  \Line(115,37)(115,50)%
  \Line(115,50)(133,53)%
  \Line(100,20)(70,40)%
  \DashLine(70,40)(115,85){1}%
  \Line(115,85)(150,70)%
  \Text(60,100)[c]{$t$}%
  \Text(125,5)[c]{$x^{ }_\perp$}%
  \Text(180,57)[c]{$L$}%
  \Text(128,30)[c]{$a$}%
  \Text(151,45)[c]{$a/v$}%
  \end{picture}
 \raise6cm\hbox{
 \includegraphics[scale=0.40, angle=-90]{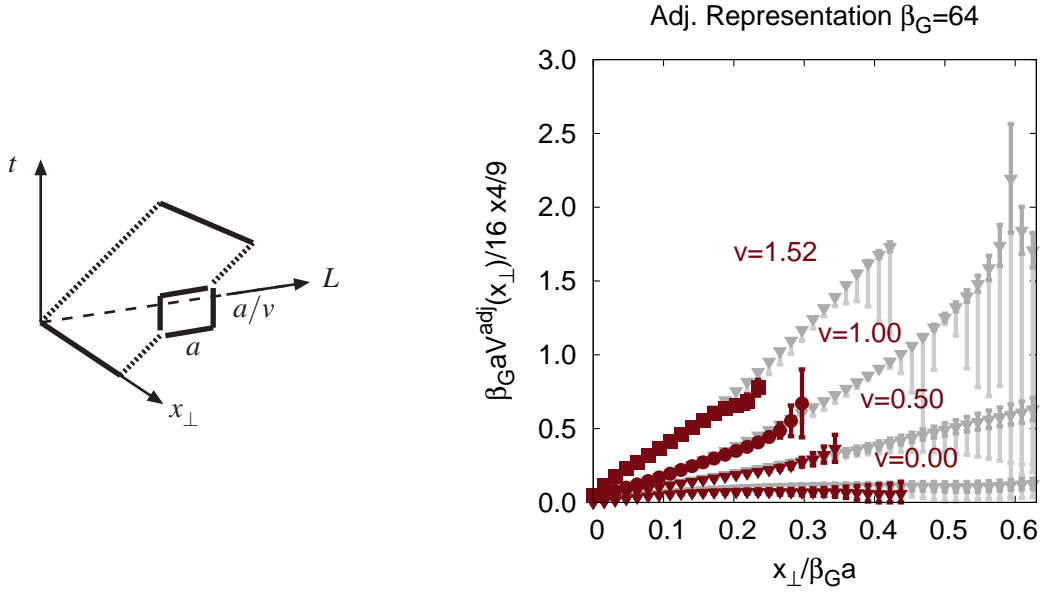}
 }
}

 \caption[a]{Left: 
A tilted lattice Wilson loop in Minkowskian spacetime.
Right: The transverse potential in the adjoint
representation (darker symbols) for different velocities at 
$\beta_\rmii{G}=64$ and $N/\beta_\rmii{G}=1.22$, 
scaled by $4/9$. 
The results for the fundamental Wilson loop~\cite{own}
at $N/\beta_\rmii{G}=1.5$ are given 
with the lighter symbols.}
\label{Fig:AdjointRep}

\end{figure}

Initial conditions are generated with the 
weight $P[U,E]\propto e^{-\beta_\rmii{G} H_\rmi{cl}} 
\Pi_{\bf x}\delta(G({\bf x}))$, by
making use of an
algorithm described in ref.~\cite{imV}. Subsequently $U$ and $E$ are
evolved in a forward Euler leap-frog scheme with temporal lattice spacing
$a_t=a/100$, based on \eqs\nr{eom1}, \nr{eom2}.  
At each integer step in time $t=n a/v$ a copy of the gauge
links is saved and a light-cone Wilson line is constructed by
appropriate averages of links above and below the actual path as shown in
\fig\ref{Fig:AdjointRep}(left). 
Monitoring the large-time behaviour of the Wilson loop, 
a transverse potential is subsequently extracted from 
the exponential damping as in \eq\nr{eq_x} (now with $L\to t$),  
\be
 V(x^{ }_\perp) \equiv - \lim_{t\to\infty} 
 \frac{\partial_t P(x^{ }_\perp,t)}{P(x^{ }_\perp,t)}
 \;. \la{damping}
\ee
The potential here is a function of $x^{ }_\perp$ as well as 
the velocity $v$, as illustrated in \fig\ref{Fig:AdjointRep}(left).
(In ref.~\cite{own} the same object was denoted by $- \im V$, 
motivated by the time evolution $\sim e^{-i E t}$.)

%
%

As an example of results that can be obtained, 
the transverse potential extracted from 
simulations with $\beta_\rmii{G}=64$ on a
$N^3=78^3$ (adjoint rep.,\ scaled with $C^{ }_\rmii{F}/C^{ }_\rmii{A} = 4/9$) 
and $N^3=96^3$ (fundamental rep.) lattice 
is shown in \fig\ref{Fig:AdjointRep}(right) as a function
of $x_\perp$. For extracting 
the damping rate from \eq\nr{damping}, a fitting range 
is needed that allows for a compromise
between signal strength on one hand and an asymptotic 
exponential behavior of the Wilson loop on the other.
Once a range is chosen, a single exponential fit is deployed to
measure the value of $V(x_\perp)$, with systematic errors estimated from 
the variation of the results when moving the fitting range to later times.

A central argument in the analysis of ref.~\cite{sch} is that the 
contribution to the thermal light-cone Wilson loop from soft (colour-electric
and colour-magnetic) gauge fields should not be sensitive to crossing
the light cone. This argument can be tested through our 
simulations: results for several $v$ are plotted 
in \fig\ref{Fig:AdjointRep}(right). 
Quantitative changes are observed as $v$ increases, but there does not
appear to be any qualitative transition in the dynamics for $v > 1$. 

To summarize, classical lattice gauge theory 
simulations support the theoretical 
arguments given in ref.~\cite{sch} and give direct physical insight of 
the behaviour of relevant observables in \linebreak
Minkowskian spacetime, 
without complications related to analytic continuation. For 
quantitative results, however, the Euclidean results of 
ref.~\cite{mp1} are to be used, because within classical
lattice gauge theory the Debye mass parameter cannot be tuned to 
a physical regularization independent value; it rather changes
rapidly with the lattice spacing. 

%
\section{How to extract the transverse collision kernel}
\la{se:inversion}

Suppose now that $V(x^{ }_\perp)$ has been 
computed non-perturbatively
at distances $x_\perp \gsim 1 / (gT)$ 
with simulations like those in ref.~\cite{mp1} and that a continuum
limit has been taken. We may then try to invert the relation 
between the potential and the transverse collision kernel, 
\eq\nr{relation}, in order to estimate $C(k_\perp)$ in the infrared domain 
$k_\perp \lsim  g T$, in which perturbation theory is slowly 
convergent~\cite{sch} (for $k_\perp \lsim g^2 T/\pi$ 
the problem becomes genuinely non-perturbative~\cite{nonpert}). 
We start by showing how the inversion can be carried out in principle.   

In the presence of an ultraviolet regularization, such as a lattice cutoff, 
the first factor on the right-hand side of \eq\nr{relation} normalizes the
potential to zero at vanishing distance. Omitting this 
overall normalization for the moment (it will be imposed in a different
fashion in a moment), 
\eq\nr{relation} can formally be inverted through a Fourier transform: 
\be
 C(k_\perp) = - \int \! {\rm d}^2\vec{x}_\perp \, 
 e^{-i \vec{k}_\perp \cdot \vec{x}_\perp} V(x_\perp)
 = 
 - 2\pi \int_0^\infty \! {\rm d}x_\perp \, x_\perp
 \, J_0(k_\perp x_\perp)
 \, V(x_\perp)
 \;, \la{step1}
\ee
where the angular integral was carried out, and $J_0$ is a Bessel function. 
The asymptotics of $J_0$, 
\be
 J_0( k_\perp x_\perp )
 \; \stackrel{ k_\perp x_\perp \gg 1 }{ \approx } \; 
 \sqrt{\frac{2}{\pi k_\perp x_\perp}} \, \cos\Bigl( 
 k_\perp x_\perp - \frac{\pi}{4} \Bigr)
 \;, 
\ee
implies however that the integral is typically not absolutely 
convergent at large $x_\perp$. For example, 
the non-perturbative asymptotics
originating from three-dimensional pure Yang-Mills 
theory obeys the string-theory predicted asymptotics~\cite{lw}  
\be
 V(x_\perp) \; \stackrel{g^2 T x_\perp  \gg 1}{\approx} \; 
 \sigma x_\perp + \mu + \frac{\gamma}{x_\perp}
 \;. \la{Vasymp}
\ee
All of these terms decay too slowly for \eq\nr{step1} to be 
absolutely integrable. 
(The coefficient $\mu$ is related to the overall normalization, 
as alluded to above.)

It is possible, however, to subtract the problematic terms and carry
out the inverse transform on a faster decaying remainder. Concretely, 
making use a dimensionally regularized Fourier transform, 
\be
 \mathcal{F}\biggl[ \frac{1}{k_\perp^\nu} \biggr]
 \;\equiv\;
 \int \! \frac{{\rm d}^{2-2\epsilon} \vec{k}_\perp}
                 {(2\pi)^{2-2\epsilon}} 
 \frac{ e^{i \vec{k}_\perp\cdot \vec{x}_\perp} }{k_\perp^\nu}
 \;=\; \frac{\Gamma(1-\epsilon - \frac{\nu}{2})}
   {\Gamma(\frac{\nu}{2})} 
   \frac{1}{2^\nu \pi^{1-\epsilon} x_\perp^{2-2\epsilon - \nu}}
 \;, 
\ee
which implies
$
 \mathcal{F}[1/k_\perp] = 1/(2\pi x_\perp)
$
as well as 
$
 \mathcal{F}[1/k_\perp^3] = - x_\perp /(2\pi) 
$, 
and tuning $\sigma,\mu, \gamma$ such that 
\be
 \lim_{x_\perp\to\infty}
 x_\perp \biggl\{ V(x_\perp) - \biggl[
  \sigma x_\perp + \mu + \frac{\gamma}{x_\perp}
 \biggr] \biggr\} = 0 
 \;,
 \la{tune}
\ee
we obtain a subtracted version of the (inverse) Fourier transform:
\be
 \frac{ C(k_\perp) }{2\pi} \; = \; 
 \frac{\sigma}{k_\perp^3} - \frac{\gamma}{k_\perp} + 
 \int_0^\infty \! {\rm d}x_\perp \, x_\perp \, J_0(k_\perp x_\perp)
 \biggl[ 
   \sigma x_\perp + \mu + \frac{\gamma}{x_\perp} - 
    V(x_\perp) 
 \biggr] 
 \;, \quad k_\perp > 0 
 \;. \la{final}
\ee
The integral here is convergent in a confining theory
(provided that the potential does not diverge too fast
at short distances, which is not the case). 

%
\section{A first numerical test}
\la{se:test}

In a practical setting, where $V(x_\perp)$ contains errors 
and is only known in a finite interval, it is not clear {\it a priori} 
whether \eq\nr{final} can yield useful results. The reason is that 
$J_0$ is oscillatory, so a kind of sign problem (significance loss)
takes place. Nevertheless the problem is less serious at
small $k_\perp$, precisely the domain of most interest, 
so it appears worthwhile to carry out a test.  

For the test, we make use of the data of ref.~\cite{mp1}.\footnote{%
 We thank Marco Panero for providing us with this data. 
 } 
As an example, we consider the so-called ``cold'' set 
($T \approx 400$~MeV) at the lattice couplings $\beta_\rmii{G} = 14,16,18$; 
these $\beta_\rmii{G}$-values are chosen as a compromise for which 
data extend both to short and large distances. The central 
values at distances $r > r_0$, where $r_0 \approx 2.2/\gE^2$
is the Sommer scale, 
are used in a $\chi^2$-minimization to determine the parameters
of \eq\nr{tune}. (For $\beta_\rmii{G} = 14$ this corresponds to the 6 largest
distances; for $\beta_\rmii{G} = 16$ to 5; for $\beta_\rmii{G} = 18$ to 4.)
Note that such a fit, at finite distances and $\beta_\rmii{G}$ and with 
the colour-electric ``decorations'' present in the Wilson loop, 
does not necessarily reproduce 
the pure Yang-Mills values~\cite{lw}, for instance
we find $\gamma> 0$. Having fixed the parameters, 
\eq\nr{final} is subsequently estimated through
\be
 \frac{ C(k_\perp) }{2\pi} \; \simeq \; 
 \frac{\sigma}{k_\perp^3} - \frac{\gamma}{k_\perp} + 
 \fr12 \sum_{i=1}^{i_\rmii{max}} 
 \bigl[ x^{ }_{\perp, i} - x^{ }_{\perp,i-1} \bigr] 
 \bigl[ \phi(x^{ }_{\perp,i}) 
 + \phi(x^{ }_{\perp,i-1})\bigr]
 \;, \la{lattice}
\ee
where $i$ numerates the distances at which data
is available, $x^{ }_{\perp,0} \equiv 0$, 
and $\phi$ denotes the integrand: 
\ba
 \phi(x_\perp) & \equiv & 
 x_\perp \, J_0(k_\perp x_\perp)
 \biggl[ 
   \sigma x_\perp + \mu + \frac{\gamma}{x_\perp} - 
    V(x_\perp) 
 \biggr] 
 \;, \quad x_\perp > 0 \;, \la{phi} \\ 
 \phi(0) & \equiv & \gamma
 \;. \la{def2}
\ea
The definition in \eq\nr{def2} originates from $J^{ }_0(0) = 1$
and the observation that $V(x_\perp)$ diverges more slowly 
than $1/x_\perp$ at short distances.
In order to produce an error band, we have generated
$\sim 100$ mock configurations
with the given central values and errors from 
ref.~\cite{mp1}, treating the errors
at various $x^{ }_{\perp,i}$ as independent from each other. 
Equation \nr{lattice} is 
evaluated for each configuration, and subsequently the central
values and their variances are determined as usual. 

\begin{figure}[t]


\centerline{%
 \epsfysize=7.8cm\epsfbox{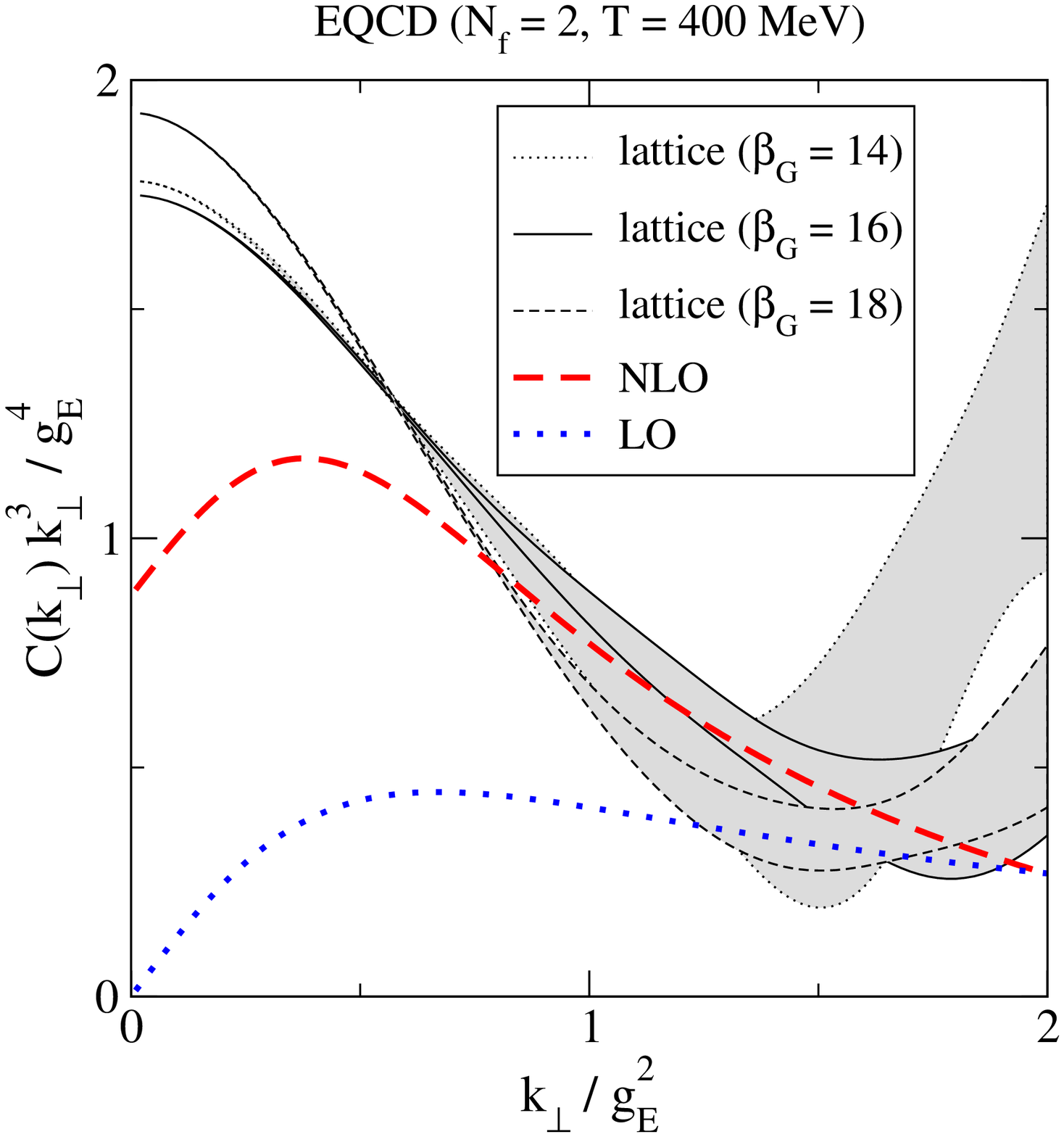}%
~~~\epsfysize=7.8cm\epsfbox{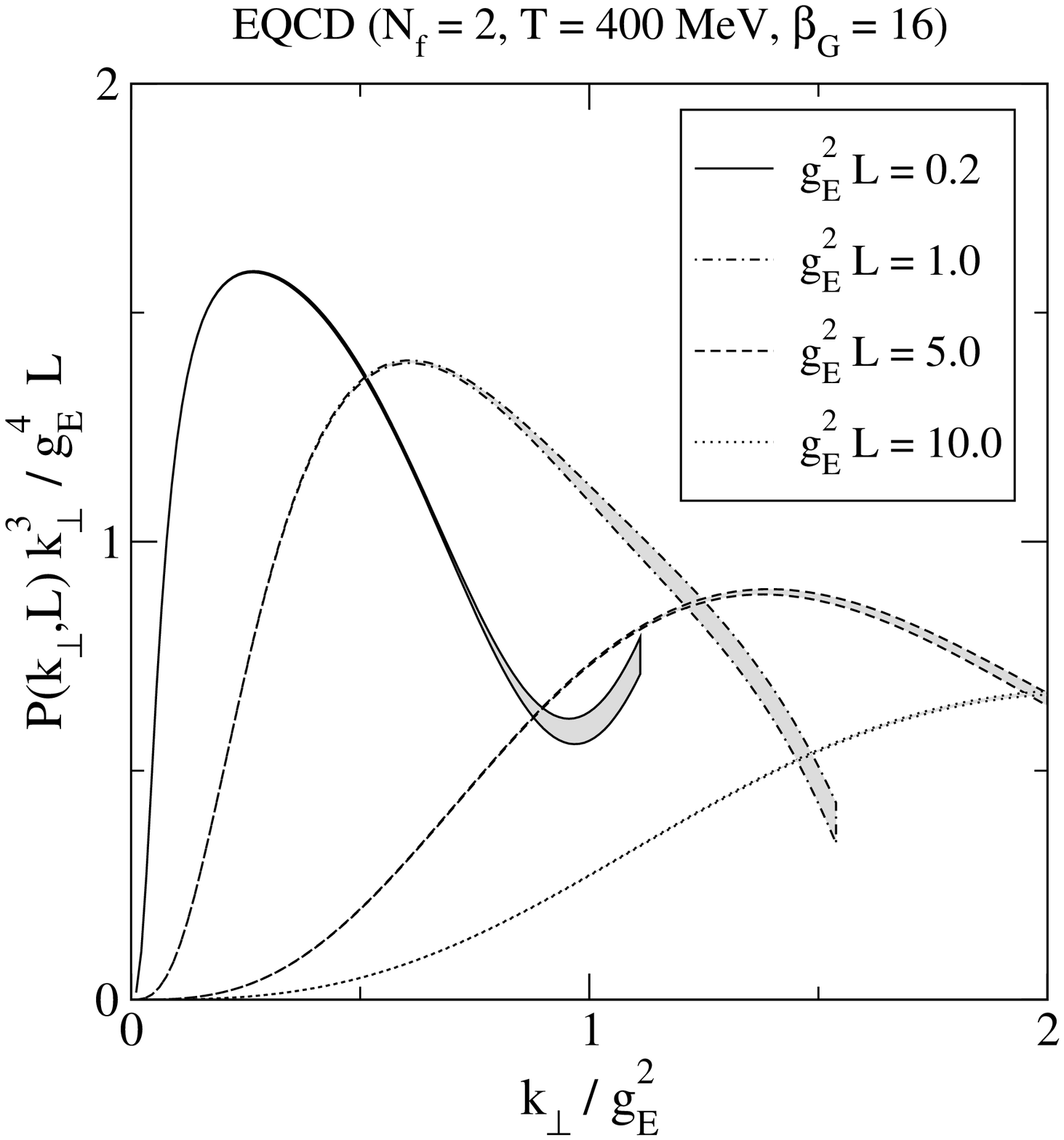}
}

\caption[a]{
Left: The transverse collision kernel extracted from \eq\nr{lattice} 
by making use of lattice data from ref.~\cite{mp1}, compared 
with the NLO result from ref.~\cite{sch}. The error
band originates from simulated statistics as described in the text. 
For numerical values of $\gE^2\sim g^2 T$ see ref.~\cite{gE2}. 
Right: A rough estimate of $P(k^{ }_\perp,L)$ from 
\eqs\nr{P_def}, \nr{eq_x}, for the data set with $\beta^{ }_\rmii{G} = 16$
(the results are again increasingly unreliable as $k^{ }_\perp$ grows).
}

\la{fig:test}
\end{figure}

The result of this procedure is shown in \fig\ref{fig:test}, 
together with the NLO result from ref.~\cite{sch}. 
A significant enhancement can be observed for 
$k_\perp < \gE^2$, where $\gE^2 \sim g^2 T$ is the effective 
coupling of the dimensionally reduced ``EQCD'' effective theory.
As $k_\perp$ increases a significance loss becomes visible; 
nevertheless, it seems conceivable that contact to perturbation
theory can eventually be made for $k_\perp \gsim \mE$.
It should be noted that at the temperature considered
the Debye mass parameter  $\mE$ (i.e.\ the electric scale) and
the gauge coupling $\gE^2$ (i.e.\ the magnetic scale) are close to
each other: ref.~\cite{mp1} 
made use of $\mE  = \sqrt{0.448306}\gE^2 \approx 0.67\gE^2 $. 

According to \fig\ref{fig:test}, $k_\perp^3 C(k^{ }_\perp)$ 
is not unlike a Gaussian at small $k^{ }_\perp$, with a height
and curvature given by the fit parameters $2\pi\sigma, 4\pi\gamma$, 
respectively.
The stability of these results with respect to adding data at smaller 
and larger distances needs, however, to be carefully investigated.

%
\section{Conclusions}

We have provided evidence that 
the remarkable proposal of ref.~\cite{sch}, namely that
purely Euclidean techniques allow to infer interesting real-time
information in a certain ``soft'' regime, appears to stand firm. 
For definite numerical conclusions it will be important to improve on 
the determination of the 
transverse collision kernel, $C(k^{ }_\perp)$, 
sketched in \fig\ref{fig:test}, by taking
the continuum limit with the data of ref.~\cite{mp1}
and exploring the systematic uncertainties 
related to \eq\nr{lattice}. 

This work was supported in part by SNF under grant 200021-140234.

%

\end{document}